\documentclass[usenatbib]{mnras}

\usepackage{newtxtext,newtxmath}
\usepackage[T1]{fontenc}
\usepackage{ae,aecompl}
\usepackage{graphicx}	
\usepackage{amsmath}	
\usepackage{amssymb}	
\usepackage{siunitx}    
\usepackage{placeins}

\usepackage{scalerel,tikz}
\usetikzlibrary{svg.path}
\definecolor{orcidlogocol}{HTML}{A6CE39}
\tikzset{orcidlogo/.pic={
 \fill[orcidlogocol] svg{M256,128c0,70.7-57.3,128-128,128C57.3,256,0,198.7,0,128C0,57.3,57.3,0,128,0C198.7,0,256,57.3,256,128z};
 \fill[white] svg{M86.3,186.2H70.9V79.1h15.4v48.4V186.2z}
 svg{M108.9,79.1h41.6c39.6,0,57,28.3,57,53.6c0,27.5-21.5,53.6-56.8,53.6h-41.8V79.1z M124.3,172.4h24.5c34.9,0,42.9-26.5,42.9-39.7c0-21.5-13.7-39.7-43.7-39.7h-23.7V172.4z}
 svg{M88.7,56.8c0,5.5-4.5,10.1-10.1,10.1c-5.6,0-10.1-4.6-10.1-10.1c0-5.6,4.5-10.1,10.1-10.1C84.2,46.7,88.7,51.3,88.7,56.8z};
}}
\newcommand\orcidicon[1]{\href{https://orcid.org/#1}{\mbox{\scalerel*{
\begin{tikzpicture}[yscale=-1,transform shape]
\pic{orcidlogo};
\end{tikzpicture}
}{|}}}}

\title[FDM Soliton Cores around SMBH]{Fuzzy Dark Matter Soliton Cores around Supermassive Black Holes}

\author[E. Y. Davies et al.]{
Elliot Y. Davies~\orcidicon{0000-0001-5996-4072}$^{1}$\thanks{E-mail: elliotdavies@princeton.edu},
{Philip Mocz~\orcidicon{0000-0001-6631-2566}$^{1}$\thanks{E-mail: pmocz@astro.princeton.edu; Einstein Fellow}}
\\
$^{1}$Department of Astrophysical Sciences, Princeton University, 4 Ivy Lane, Princeton, NJ, 08544, USA\\
}

\date{Accepted XXX. Received YYY; in original form ZZZ}

\pubyear{2019}

\begin{document}

%
%
%
%
%
%
%
%

\label{firstpage}
\pagerange{\pageref{firstpage}--\pageref{lastpage}}
\maketitle

\begin{abstract}

We explore the effect of a supermassive black hole (SMBH) on the density profile of a fuzzy dark matter (FDM) soliton core at the centre of a dark matter halo. We numerically solve the Schr\"odinger-Poisson equations, treating the black hole as a gravitational point mass, and demonstrate that this additional perturbing term has a `squeezing' effect on the soliton density profile, decreasing the core radius and increasing the central density. In the limit of large black hole mass, the solution approaches one akin to the hydrogen atom, with radius inversely proportional to the black hole mass. By applying our analysis to two specific galaxies (M87 and the Milky Way) and pairing it with known observational limits on the amount of centrally concentrated dark matter, we obtain a constraint on the FDM particle mass, finding that the range $10^{-22.12}\:\si{eV}  \lesssim m \lesssim 10^{-22.06}\:\si{eV}$ should be forbidden (taking into account additional factors concerning the life-time of the soliton in the vicinity of a black hole). Improved observational mass measurements of the black hole and total enclosed masses will significantly extend the lower-bound on the excluded FDM mass region, while self-consistent theoretical modeling of the soliton--black hole system can extend the upper-bound.

\end{abstract}




\numberwithin{equation}{section}

\section{Introduction}


The exact particle nature of dark matter (DM) remains one of physics and astronomy's biggest mysteries, and while the cold dark matter (CDM) model has had significant success in explaining large-scale structure of the Universe, this widely popular model does face open challenges on scales on the order of galaxy size \citep{weinberg2015cold, bullock2017small}. Most notably there is the (1) `cusp-core' problem \citep{moore1994evidence, flores1994observational}, where the density profile of CDM halos is expected to be cuspy ($\rho \sim r^{-1}$) from simulations instead of the observed core-like ($\rho \sim r^{0}$) profile, as well as the (2) `missing satellites' problem \citep{moore1999dark, klypin1999missing}, whereby CDM predicts a larger number of high-luminosity satellite galaxies than is observed, and the related (3) `too-big-to-fail' problem concerning the internal dynamics of bright satellites \citep{boylan2011too}. Despite the prevailing attitude that CDM is the apex DM theory, the current lack of evidence for any CDM particle, such as the WIMP (or \textit{weakly interacting massive particle} with $m \approx \mathcal{O}(\si{GeV})$), is a compelling reason to look to a promising alternative: DM as an ultralight scalar field with spin-0 \citep{hu2000fuzzy, goodman2000repulsive}. Often called \textit{fuzzy dark matter} (FDM), this scalar field is assumed to have a particle mass of $\sim 10^{-22}\:\si{ eV}$ and a de Broglie wavelength of $\lambda = 1.2 \left( \frac{m}{10^{-22}~\si{eV}}\right)\left(\frac{100~\si{km}~\si{s}^{-1}}{v}\right)\:\si{kpc}$ which gives rise to a quantum pressure that smooths cosmological small-scale structure and stabilises the centres of DM halos, while making the same large-scale predictions as CDM (e.g. \citealt{hu2000fuzzy,mocz2018schrodinger,mocz2019first,mocz2019galaxy}). \\


FDM predicts a unique stable structure at the centre of DM halos -- the \textit{soliton} core -- where the quantum pressure prevents collapse against self-gravity.
The predicted existence of solitons can give us useful insight into the utility of the dark matter theory when compared with observational constraints. Solitons may largely dictate bulge dynamics of large galaxies, and even constitute almost all of the halo mass in smaller dwarf galaxies in the FDM model. \\

Most galaxies are also known to harbour a supermassive black hole (SMBH) of millions of solar masses at their centre that can dominate the central mass content of the galaxy \citep{ferrarese2000fundamental,gebhardt2000relationship}. Thus the impact of the SMBH on the density profile of the soliton merits study, and the exploration of this is the primary goal of our work. 
Scalar fields around black holes have been studied in a fully general relativistic context \citep{barranco2012schwarzschild,barranco2017self,avilez2018possibility,hui2019black}, and seen to give rise to long-lasting quasi-stable `scalar wigs' surrounding the vicinity of `no-hair' black holes. 
Here we are interested in developing simple theory for the soliton profile far from the Schwarzschild radius and take a simplified approach in a regime where the SMBH can be treated as a point mass, applicable to FDM cosmology.
Our theory is also applicable to other baryonic perturbing forces on the soliton (see e.g. \cite{veltmaat2019baryon}).
We point out that similar approaches to studying the soliton plus black hole system have also been recently undertaken in \cite{chavanis2019mass,bar2019looking}. \\

An understanding of the effect of SMBH on soliton cores is a vital ingredient in the study of DM density profiles at galactic cores. 
Central DM density profiles can be paired with known observational constraints of the central density to provide constraints on the FDM model itself. Solitons are a key smoking-gun prediction of the FDM model that can be used to rule it out or confirm it. \\

There has been significant effort in recent years to place various astrophysical constraints on the FDM particle mass. 
Constraints come from a variety of systems and scales (from the cosmic microwave background ($\gg 10~\si{Mpc}$) to the centres of galaxies ($\sim 1~\si{pc}$)).
These include:
(1) using stellar velocity dispersion to fit the Milky Way's dwarf spheroidal galaxies with a soliton core assuming the systems are dark-matter dominated \citep{marsh2015axion,gonzalez2017unbiased,broadhurst2019ghostly,safarzadeh2019ultra},
(2) fitting ultra-faint dwarfs with soliton-cored halo models \citep{safarzadeh2019ultra},
(3) using the abundance of Milky Way satellites \citep{nadler2019constraints},
(4) analysing the Lyman-$\alpha$ forest as a tracer of dark matter structure \citep{kobayashi2017lyman,armengaud2017constraining,irvsivc2017first,nori2018lyman},
(5) placing constraints from CMB lensing \citep{hlovzek2018using},
(6) calculating dynamical heating on the Milky Way's stellar disc from FDM substructure including interference pattern fluctuations \citep{church2019heating},
(7) calculating the impact of FDM fluctuations on stellar streams that formed from disrupted globular clusters in the Milky Way \citep{amorisco2018first},
(8) calculating the impact of soliton cores on galactic nuclei assuming they mimic SMBHs \citep{desjacques2019axion},
(9) modeling of ultra-diffuse galaxies \citep{2019arXiv190510373W},
(10) effects of FDM on the iso-curvature component in the DM power spectrum \citep{feix2019isocurvature}, and 
(11) black-hole superradiance \citep{cardoso2018constraining}.
There has also been a recent claim for the Milky Way that a central solitonic core of mass $\simeq10^9~\si{M_{\odot}}$ and particle mass $m\simeq 10^{-22}~{\rm eV}$ is observationally supported by the central motion of bulge stars \citep{2018arXiv180708153D}.
\\

The FDM mass constraints generally fall into two camps that are in moderate tension.
Dwarf spheroidal galaxies are typically well fit by large, low-density dark matter cores, such as the soliton cores predicted by FDM theory with a particle mass of $m\simeq 10^{-22}~\si{eV}$. But many other investigations favor $m>{\rm few}\times 10^{-22}~\si{eV}$, where the wave effects of FDM are smaller.
Thus, FDM faces a sort of \textit{Catch-22} problem that warm dark matter theory faces to a larger extent \citep{maccio2012cores}.
We point out that the mild evidence in favor of soliton cores in dwarf galaxies from stellar velocities is not a strict constraint on the particle mass in the same sense that CDM cannot be considered ruled out by these measurements.
But if indeed the FDM particle mass is small, then the lack of consistency with other measurements may possibly be due to systematic biases from certain model-dependent assumptions that are used when testing the FDM model on different scales. 
For example, Lyman-$\alpha$ constraints use models that only consider the cut in the initial power spectrum of dark matter and ignore wave effects like interference pattern fluctuations of the FDM field which can add additional small-scale power.
FDM simulations have also generally ignored the feedback of baryons on the dark matter, which is important in the halo centers of CDM.
{\it Gaia} will also improve the phase-space sampling of dwarf satellites, which will better constrain dark matter potentials \citep{2019arXiv190708841L}.
We therefore take the view that it is important to obtain as many independent constraints as possible, over a range of physical scales, to verify existing constraints. \\

In this work, we are able to place a new independent constraint on the FDM particle mass, that comes from small scale ($<10~\si{pc}$) data, namely upper limits on the amount of dark matter concentrated around the SMBHs of the Milky Way and M87 (which recently had its black hole imaged; \citealt{event2019first}) that come from dynamical measurements \citep{gebhardt2011black,do2019relativistic}.
We highlight our findings here, namely that
we rule out an FDM particle of mass in the range.
$10^{-22.12}\:\si{eV}  \lesssim m \lesssim 10^{-22.06}\:\si{eV}$ (from M87*) and obtain no constraint from Sgr A*.
Recently, \cite{bar2019looking} has also looked at placing similar constraints using data from Milky Way and M87.
Our analysis is strictly more conservative in its assumptions (which we lay out in the manuscript), takes a simplified approach, and uses newer dynamical data, e.g. \cite{do2019relativistic} which constrains dark matter concentration in the vicinity of Sgr A* by an order of magnitude compared to previous measurements \citep{boehle2016improved}, and highlights avenues to improve the results in the near future.
Our results are consistent with the findings of \cite{bar2019looking}:
$m \lesssim 4.0\times 10^{-22}\:\si{eV}$ (from M87*)
$2.0\times10^{-20}\:\si{eV}  \lesssim m \lesssim 8.0\times10^{-19}\:\si{eV}$ (from Sgr A*),
highlighting the robustness of our method, but also reducing the span of the ruled-out parameter space due to relaxed assumptions. \\

Our paper is organised as follows. In section~\ref{section2} we describe the necessary mathematical background of the FDM system with a point mass black hole perturber and solve for the ground state numerically. In section~\ref{section3} we predict soliton core density profiles in halos with SMBHs, as a function of halo mass and FDM particle mass. Section~\ref{section4} details our method of obtaining our own constraint for the FDM particle mass from the Milky Way and M87, and our concluding remarks are laid out in section~\ref{section5}. \\

\section{Schr{\"o}dinger-Poisson system with point mass perturber}\label{section2}

If DM is comprised of ultralight FDM particles, the occupation numbers in galactic halos are so high \citep{tremaine1979dynamical} that the dark matter behaves as a classical field obeying the wave equation. The scalar field is minimally coupled to gravity and interacts only gravitationally with baryonic matter. The system obeys the Schr{\"o}dinger-Poisson (SP) equations \citep{moroz1998spherically, hu2000fuzzy, bahrami2014schrodinger,mocz2015numerical}. \\

We consider the SP equations for FDM described by the wavefunction $\psi$  with an added black hole perturber $V_{\bullet}$:

\begin{equation}
i\hbar\psi_t = \left( \frac{-\hbar^2}{2m}\nabla^2 + m(V+V_{\bullet})\right)\psi    
\end{equation}

where the self-potential $V$ of the wavefunction obeys Poisson's equation

\begin{equation}
    \nabla^2 V = 4\pi G |\psi|^2
\end{equation}

and the black hole perturber is a point mass potential

\begin{equation}
    V_{\bullet} = - \frac{GM_{\bullet}}{r}  \,.
\end{equation}

In our notation $M_{\bullet}$ is the mass of the black hole and $m$ is the mass of the FDM particle.
The wavefunction is normalised such that the physical density is $\rho=|\psi|^2$.
 $G$ is the gravitational constant.\\

To construct spherical steady-state solutions, we make the ansatz that $\psi(r,t) = e^{-i\gamma t/ \hbar}\phi(r)$, where $\phi(r)=\sqrt{\rho(r)}$. It is convenient to define a number of dimensionless variables to simplify numerical calculation:

\begin{gather}
\hat{\phi} =  \frac{\hbar\sqrt{4\pi G}}{mc^2} \phi \\
\hat{r} =  \frac{mc}{\hbar} r \\
\hat{t} =  \frac{mc^2}{\hbar} t \\ 
\hat{V}_{\rm self} =  \frac{1}{c^2} V_{\rm self} \\
\hat{V}_{\bullet} =  \frac{1}{c^2} V_{\bullet} \\
\hat{\gamma} =  \frac{1}{mc^2} \gamma
\end{gather}

so we can rewrite the SP system as

\begin{gather}\label{SPhydrogen}
\hat{\gamma} \hat{\phi} = \left( -\frac{1}{2}\hat{\nabla}^2 + \hat{V}_{\rm self} +\hat{V}_{\bullet}  \right) \hat{\phi} \\
\hat{\nabla}^2 \hat{V}_{\rm self}  = \hat{\phi}^2 \\
\hat{V}_{\bullet}  = -\frac{\hat{\alpha}}{\hat{r}}
\end{gather}

where we define the parameter $\hat{\alpha} \equiv \frac{GM_{\bullet}m}{\hbar c}$, which sets the dimensionless mass of the black hole. Notice that, without the self-potential, equation~\ref{SPhydrogen} is equivalent to the Schr\"odinger equation for the hydrogen atom, which has ground state solution $\phi(r) \propto \exp(-\hat{r}\hat{\alpha})$ with $\hat{\gamma} = -\hat{\alpha}^2$/2. In dimensionful units, the length

\begin{equation}\label{bohrradius}
    \alpha^{-1} = \frac{(\hbar/m)^2}{GM_{\bullet}}
\end{equation}

is equivalent to the Bohr radius of the hydrogen atom. Therefore in the limit of a dominant black hole perturber, it is the mass of the black hole that sets the radius of the soliton core. \\

Following the method of \citet{guzman2004evolution} we arrive at the the following set of ordinary differential equations, albeit with the additional black hole term on the RHS:

\begin{align}\label{SN1}
(\hat{r}\hat{\phi})_{\hat{r}\hat{r}} &= 2\hat{r}\left(\hat{V} - \hat{\gamma}-\frac{\hat{\alpha}}{\hat{r}}\right),  \\
(\hat{r}\hat{V})_{\hat{r}\hat{r}} &= \hat{r}\hat{\phi}^2.\label{SN2}
\end{align}

\begin{figure}
    \centering
    \includegraphics[width=0.47\textwidth]{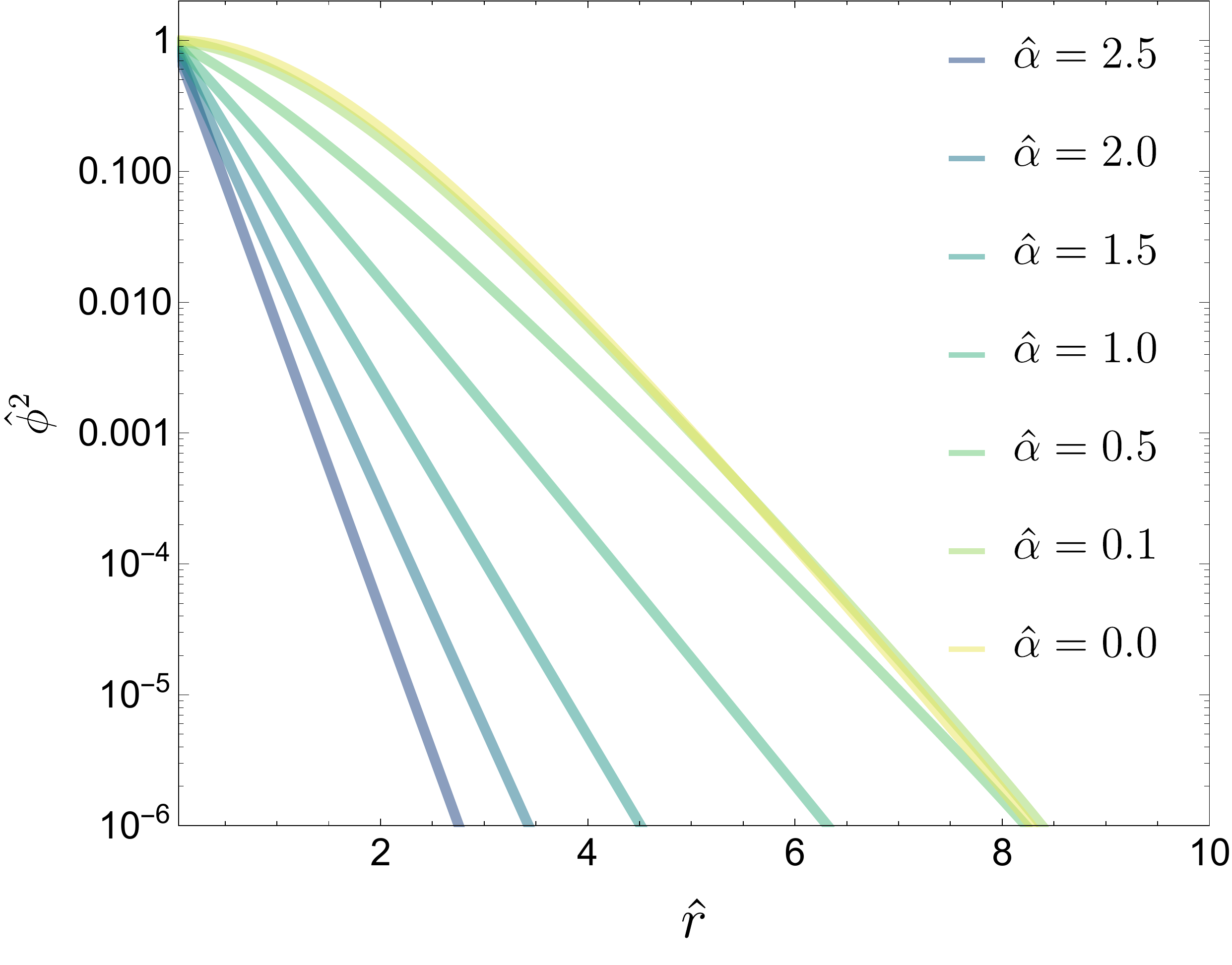}
    \caption{Ground state numerical solutions of the SP system in dimensionless units for $\hat{\alpha} = 0$ to $2.5$, using boundary condition $\hat{\phi}(\hat{r} = 0) = 1$. In the presence of a massive black hole, the soliton core approaches an exponential profile solution (akin to the hydrogen atom ground state).
    }
    \label{fig:dimlesssol}
\end{figure}

Using arbitrary normalisation $\hat{\phi}(\hat{r} = 0) = 1$ and $\hat{V}(\hat{r} = 0) = 0$, and using boundary conditions $\hat{\phi}'(\hat{r} = 0) = 0$, $\hat{V}'(\hat{r} = 0) = 0$, and $\hat{\phi}(\hat{r} = \infty) = 0$, the  SP system turns into an eigenvalue problem for $\phi(r)$ with a unique set of values of $\hat{\gamma}$ for any unique value of $\hat{\alpha}$ for which the above conditions are fulfilled. Only the 0-node $\hat{\gamma}$ (henceforth $\hat{\gamma}_0$) is stable \citep{lora2014sextans} so we only take the lowest value of $\hat{\gamma}$ for any physical model. We solve the system of ordinary differential equations using a shooting method in \textsc{Mathematica}. The method solves for the solution starting from $\hat{r}=0$, 
and tries to reach the asymptotic boundary condition $\hat{\phi}(\hat{r} = \infty) = 0$.
For this a value of $\hat{\gamma}$ needs to be assumed (guessed), and only quantised values will satisfy the asymptotic boundary conditions. We use a standard root finding technique to find the smallest $\hat{\gamma}_0$ that yields a solution that satisfies the boundary condition to yield the $0$-node (ground state) solution.
Figure~\ref{fig:dimlesssol} shows the dimensionless solutions we obtained for various values of $\hat{\alpha}$. In the limit of large $\alpha$ (i.e., large black hole mass) an exponential solution of the hydrogen atom is seen to be recovered. \\

For each unique solution, the total mass of the soliton is 

\begin{equation}\label{massestimate}
    M = \int_0^{\infty}|\psi|^2 4\pi r^2 dr = \frac{\hbar c}{G m} \int_0^{\infty}\hat{\phi}^2\hat{r}^2 d\hat{r} \,.
\end{equation}

As with the previous variables we define a dimensionless mass:

\begin{equation}\label{dimensionlessmass}
    \hat{M} \equiv \frac{GMm}{\hbar c} = \int_0^{\infty}\hat{\phi}^2\hat{r}^2 d\hat{r} \,.
\end{equation}

A useful quantity to consider is the ratio of black hole mass to soliton mass which we will denote as

\begin{equation}
    \Xi \equiv \hat{\alpha} / \hat{M} = M_{\bullet}/M 
\end{equation} 

throughout this paper.  \\

The SP system has the scaling symmetry:

\begin{gather}
    \hat{\phi} \rightarrow \lambda\hat{\phi} \\
    \hat{V} \rightarrow \lambda\hat{V} \\
    \hat{V_{\bullet}} \rightarrow \lambda\hat{V_{\bullet}} \\
    \hat{\alpha} \rightarrow \lambda^{1/2}\hat{\alpha} \\
    \hat{\gamma} \rightarrow \lambda\hat{\gamma} \\
    \hat{r} \rightarrow \lambda^{-1/2}\hat{r} \\
    \hat{M} \rightarrow \lambda^{1/2}\hat{M} \label{masslambda}
\end{gather}

which can be used to transform our dimensionless numerical solution to physical units of a given soliton mass $M$. \\

The mass ratio $\Xi$ is invariant under scaling and is the primary parameter setting the shape of the soliton. In the limit that $\Xi \gg 1$, we recover the exponential solution analogous to that of the hydrogen atom with the aforementioned black hole Bohr radius (equation~\ref{bohrradius}). In the case of no black hole ($\Xi = 0$), our numerical solutions confirmed the analytic fit to the density of a soliton done by \citet{schive2014cosmic}:

\begin{equation}\label{coreradius}
    \rho_c(r) \simeq \frac{0.019\times(m/10^{-22}~\si{eV})^{-2}(r_c/\si{kpc})^{-4}}{\left[1 + 0.091\times(r/r_c)^2\right]^8}~\si{M_{\odot}{\rm pc}^{-3}}
\end{equation}

where $r_c$ is the radius of the soliton core. The mass of such a soliton is inversely proportional to its core radius, as the soliton obeys

\begin{equation}\label{eqn:mr}
    M \times r_c \simeq 2.2\times10^8 (m/10^{-22}~\si{eV})^{-2}~\si{M_{\odot}}{\rm kpc} \,.
\end{equation}

\section{Applications to halos with a central soliton and SMBH}\label{section3}

Our primary aim in this section is to model the density profile of the soliton for a given halo mass (and as a function of the FDM particle mass), using our numerical solutions from section~\ref{section2}.
We assume that in a cosmological context the halo and soliton core form on a free-fall time and establish a thermodynamic equilibrium which sets the soliton mass. Then, the SMBH, which grows later and may come to dominate the central mass of the halo, is a perturbing force on the centre of the system.
We compare the solution both with and without the presence of a SMBH.\\

\subsection{Soliton-halo mass relation}

The soliton cores in FDM halos are known to scale with the halo mass.
From cosmological dark matter-only simulations of FDM conducted in \citet{schive2014understanding} it was observed that soliton mass is related to halo mass by 

\begin{equation}
    M \propto a^{-1/2}M_{\text{halo}}^{1/3}
\end{equation}

where $a$ is the cosmological scale factor.
We take the scaling relation at the present age of the Universe ($z = 0$, $a=1$). The precise scaling is

\begin{equation}\label{solitonhalo}
    M = 1.25\times 10^9 \left(\frac{M_{\text{halo}}}{10^{12}~\si{M_{\odot}}}\right)^{1/3} \left(\frac{m}{10^{-22}~\si{eV}}\right)^{-1}~\si{M_{\odot}} \,,
\end{equation}

The physical meaning of the scaling relation is that the size of the soliton matches the de Broglie wavelength of the velocity dispersion $\sigma$ of the halo.
The relation is predicted from a thermodynamic equilibrium argument from potential theory \citep{binney2011galactic}
by equating $GM/R\sim GM_{\rm halo}/R_{\rm halo} \sim \sigma^2$ = gravitational `temperature' between the soliton and the halo.
We comment that in non-cosmological, fully-virialized settings \citep{mocz2017galaxy} the soliton mass may be higher, however we take the \cite{schive2014understanding} cosmological relation which is more conservative for our purposes. Solitons at halo centers may also exhibit strong quasi-normal oscillations \citep{veltmaat2018formation}, an effect that we do not consider here.
\\

\begin{figure*}
  \centerline{\includegraphics[width=0.97\textwidth]{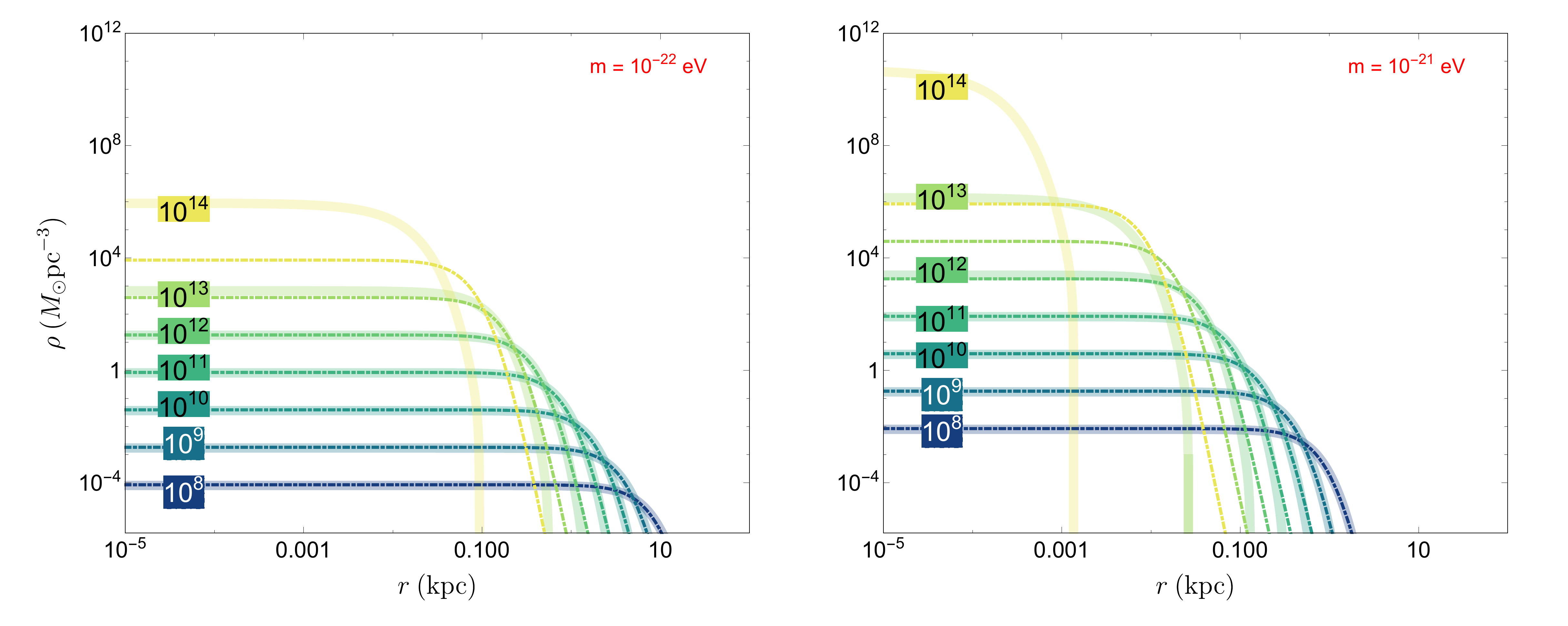}}
  \caption{Numerical solutions of the SP system for halo masses from $10^8~\si{M\odot}$ up to $10^{14}~\si{M\odot}$ for a FDM particle mass of $m = 10^{-22}~{\rm eV}$ (left) and $m = 10^{-21}~{\rm eV}$ (right). The thin, dot-dashed lines correspond to the $\hat{\alpha} = 0$ (without SMBH) solutions and the thicker, slightly opaque lines correspond to $\hat{\alpha}>0$ (with SMBH) solutions.  Dimensionless numerical solutions $\phi(r)$ have been appropriately scaled to a dimension-full solutions $\rho$ by comparison of core densities of numerical $\hat{\alpha} = 0$ with the core densities of analytic solutions obtained from equation~\ref{coreradius}.}
  \label{fig:haloprofiles}
\end{figure*}

\subsection{Black hole-halo mass relation}

Almost all galaxies (except small dwarfs) have a SMBH at their centres.
Observationally, the SMBH mass is known to correlate tightly with the velocity dispersion $\sigma$ of the host-galaxy bulge, the so-called `$M_{\bullet}$--$\sigma$' relation  \citep{sun2013refining,kormendy2013coevolution}.
As a result, the SMBH mass is also correlated with global halo properties, such as the total halo mass \citep{ferrarese2002beyond,bandara2009relationship}.
There is of course scatter in the relation (e.g., see the MASSIVE survey results on the most massive SMBHs; \citealt{ma2014massive}).
But it is reasonable for us to assume a fiducial relationship between SMBH and halo mass to estimate how solitons in halos of different masses would be typically affected. \\

We take as our fiducial relationship the expression from \citet{bandara2009relationship}:

\begin{equation} \label{blackholehalo}
    \log_{10}({M_{\bullet}/M_{\odot}}) = 8.18 + 1.55 \times \left[\log_{10}({M_{\text{halo}}}/M_{\odot}) - 13.0\right] \,.
\end{equation}

This relationship was derived from subset of the galaxy-scale strong gravitational lenses from the Sloan Lens ACS (SLACS) Survey \citep{bolton2006sloan}. The $M_{\bullet}$--$M_{\text{halo}}$ relationship was obtained by estimating the masses of the SMBHs in their sample by combining the $M_{\bullet}$--$\sigma$ relation from \citet{gultekin2009m} and the $M_{\bullet}$--$n$ (where $n$ is the Sersic index) relationship from \citet{graham2007log}. \\

\subsection{Density profiles}

Combining equations~\ref{solitonhalo} and \ref{blackholehalo} allows us to predict the mass ratio of the SMBH to the soliton core, $\Xi \equiv M_{\bullet} / M$, for a given halo mass $M_{\text{halo}}$. As we have stated, we assume that the soliton-halo mass relation still holds in the presence of a SMBH which we treat as a perturber to an established soliton-halo system, and that the gravitational field of the black hole will reshape the soliton density profile. 
We discuss alternative assumptions for the soliton mass with the additional presence of the SMBH, and their effect on our calculations, in section~\ref{sec:assump}. \\

We obtain the soliton density profile for the given halo mass as follows.  \\

In our numerical solutions we had set the arbitrary boundary condition
$\hat{\phi}(0)=1$, and also set the black hole mass through the dimensionless parameter $\hat{\alpha}$.
This yielded a solution with a consequently-deduced mass ratio $\hat{\alpha} / \hat{M}$.
So here we find $\hat{\alpha}$ that yields a solution that matches the predicted mass ratio $\Xi$ for our given halo of interest. We then convert the dimensionless density profile to dimensionful units, and use the $\lambda$-scaling symmetry relations described in section~\ref{section2} to convert the soliton solution so that its total mass matches that predicted by the soliton-halo mass relation (equation~\ref{solitonhalo}).
This is done by comparing the mass of the unscaled numerical solution (\ref{massestimate}) to equation~\ref{solitonhalo} for the given halo mass. Then using the found $\lambda$ from equation~\ref{masslambda}, we can scale the numerical solution. 
Appendix~\ref{append} lists some numerical values of $\hat{\alpha}, \: \hat{\gamma}_0$, $\hat{M}$ that were found in the process for various halo masses. \\

For illustrative purposes we calculate the FDM core profiles of halo masses ranging from $10^8$M$_{\odot}$ to $10^{14}$M$_{\odot}$ in order to visualise results for sensible range of galaxy halo masses. We consider the cases of a FDM particle mass of $m = 10^{-22}~\si{eV}$ and  $m=10^{-21}~\si{eV}$.
 Figure~\ref{fig:haloprofiles} shows the core profiles for the various halo masses. It can be seen that the effect of black hole perturber can only be noticed for $M_{\text{halo}} \geq 10^{13}~\si{M_{\odot}}$ for the smaller value $m$ and for $M_{\text{halo}} \geq 10^{12}~\si{M_{\odot}}$ for the larger value of $m$.
 That is, the SMBHs are most effective at modifying the halo cores for higher halo masses, and larger FDM particle masses.  The effect is that of `squeezing' the density profile inwards to have a decreased core radius but increased central density. 
 When the black hole in unimportant, the core radius scales inversely with the soliton mass (equation~\ref{eqn:mr}).
 But when the black hole dominates, the soliton size now scales inversely with the black hole mass (equation~\ref{bohrradius}), which can make solitons orders of magnitude more compact in certain cases. \\
 \\

\subsection{Soliton accretion time}

An important question to ask is whether the soliton can survive given the presence of a SMBH at its centre or if it would be accreted. 
By the `no-hair' theorem, the soliton cannot survive forever.
An approximate estimate for the soliton accretion timescale by the black hole is given in
\citet{hui2017ultralight}, which uses the absorption timescale for plane-waves by a Schwarzschild black hole \citep{unruh1976absorption}.
The estimate assumes a non-rotating black hole of mass $M_{\bullet}$ traveling at speed $v$ through a uniform scalar field of mass $m$ and density $\rho$, which accrete mass at a rate 

\begin{equation}
    \frac{dM_{\bullet}}{dt} = \frac{32\pi^2(GM_{\bullet})^3m\rho}{\hbar c^3 v \left[1 - \exp(-\xi) \right]}
\end{equation}

where $\xi = 19.07 M_{\bullet}/M$, and $\rho$ is estimated to be the central density of the soliton and $v$ to be the virial velocity of the soliton. 

We can define the time in which the soliton will be accreted by the black hole as 

\begin{equation}\label{accrete}
    t_{\text{acc}} = \frac{M}{(dM_{\bullet}/dt)} \,.
\end{equation}

However, a more careful calculation, relevant for solitons around black holes, is performed in \cite{barranco2017self}, 
which find zero angular momentum $\ell=0$ solutions can form quasi-bound states with a longer accretion time.
The decay/absorption of the quasibound states is due to the tunneling of the scalar field through the potential barrier towards
the black hole horizon. 
According to \cite{barranco2017self}, the scalar field is quasibound when 
$\frac{M_{\bullet}}{10^8 M_\odot} \frac{m}{10^{-22}~{\rm eV}} \ll 1.3\times 10^4$.
In the case that the potential of the black hole dominates over the scalar field ($\Xi=M_{\bullet}/M\gtrsim 1$), 
the accretion timescale is given by \citep{barranco2011black,barranco2017self}:
\begin{equation}\label{accrete2}
    t_{\text{acc}} =  5.6\times 10^{18} \left( \frac{M_{\bullet}}{10^8 M_\odot}  \right)^{-5} \left( \frac{m}{10^{-22}~{\rm eV}} \right)^{-6}~{\rm yr} \,.
\end{equation}



For our purposes, we will use the second estimate, which actually gives comparable but slightly longer timescales than the simple estimate using plane-waves.
We point out that there is room for improvement in the theoretical time-dependent modeling of this problem.
There may actually be a stabilizing effect that makes the soliton cores even longer-lived, due to the solitons becoming less compact as they lose mass
(equation~\ref{eqn:mr}). In other contexts, such as tidal disruption of solitons, this unusual `equation of state' for the soliton leads to runaway effects as the soliton loses mass and expands more of itself to beyond the tidal radius \citep{du2018tidal}.

\begin{figure}
    \centering
    \includegraphics[width=0.47\textwidth]{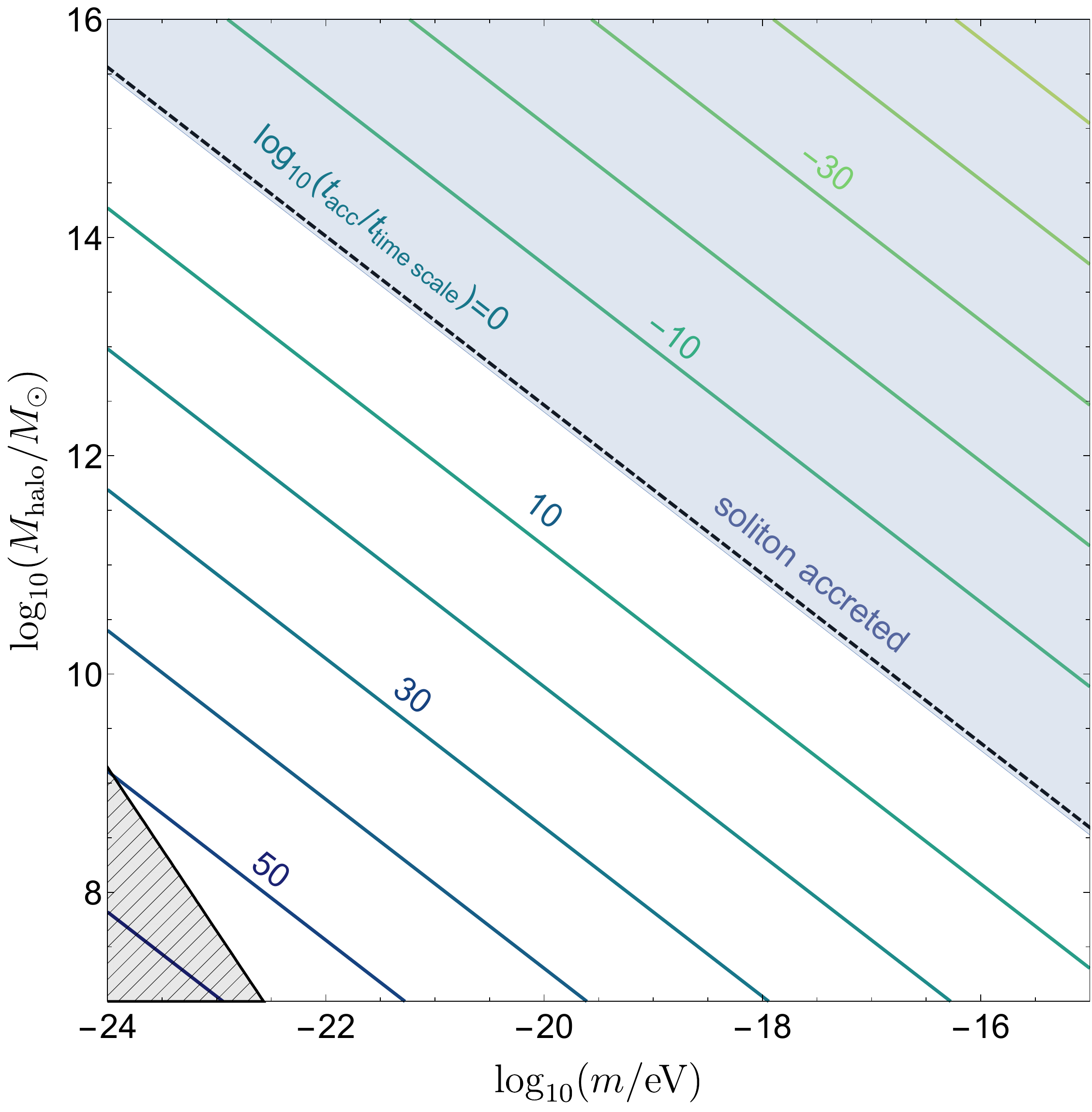}
    \caption{Contour plot of the log of the accretion time (in years) normalised by the age of the Universe, as a function of $\log_{10}(M_{\text{halo}})$ and $\log_{10}(m)$. The dotted line represents the contour with a value of our relevant timescale: $10^{10}~\si{yr}$. The hatched region in the bottom left corner corresponds to where the soliton mass is below $M_{\text{min}}$, implying that no soliton/halos can exist for these values in FDM cosmology theory.}
    \label{fig:acctime}
\end{figure}

The accretion time needs to be taken into consideration as we attempt to place constraint on the FDM particle mass: if for some given $m$, the accretion time is less than some relevant characteristic timescale, say the age of the Universe ($13.7\times10^{9}~{\rm yr}$) or the re-condensation time of the soliton (section~\ref{sec:recond}), then the soliton would have been accreted by the black hole and would not be present in the halo. For our purposes for modeling M87 and the Milky Way in subsequent sections, we found re-condensation of the soliton is not relevant (section~\ref{sec:recond}), and the characteristic timescale we found most appropriate is $10^{10}$ years, an estimate for how long the black hole and halo may have existed in the Universe with masses close to their present day values (see, e.g. \cite{sanchez2018preferential} and \cite{colin2016cosmological}). For a contour plot of accretion time across a range of FDM particle masses and halo masses see Figure~\ref{fig:acctime}, where we have indicated reference timescale as the dotted black line.
In the figure, we have again assumed the soliton-halo mass relation (equation~\ref{blackholehalo}) and the black hole-halo mass relation (equation~\ref{solitonhalo}) to associate black hole and soliton masses for the given halo mass.  \\

It is important to note that this accretion time is a conservative (lower limit) estimate for the life-time of the soliton as it assumes that the black hole mass $M_{\bullet}$ had its present day value over the entire history of the Universe. The estimated life-time of the soliton would be longer if the black hole were less massive in its past. It is known that many black holes likely grow exponentially over time on scales as fast as the Salpeter time, $t_{\rm Sal} = 5 \times 10^7$~yr, (i.e., Eddington-limited accretion) in order to build up the cosmic SMBH mass function inferred from the X-ray background  \citep{soƚtan1982masses}, which would mean the soliton life-times could longer than our estimates.
Realistic simulations of the SMBH in the Milky Way, incorporating secular accretion and mergers, suggest that Sgr A* that the last mass doubling of the black hole mass may have happened on timescales of $10^{10}$ year \citep{sanchez2018preferential}, which is the conservative estimate we consider.
\\

We point out that in a cosmological setting, there is a minimum mass soliton that can exist which has to have central density about $\sim 200\times$ the critical density of the Universe \citep{hui2017ultralight}, which is given by:
\begin{equation}
    M_{\rm min} \simeq 1.4\times 10^7  \left( \frac{m}{10^{-22}~{\rm eV}}\right)^{-3/2}  M_\odot \, ,
\end{equation}
as well as a maximum mass \citep{hui2017ultralight} above which the soliton collapses directly into a black hole:
\begin{equation}
    M_{\rm max} \simeq 8.46\times 10^{11}  \left( \frac{m}{10^{-22}~{\rm eV}}\right)^{-1}  M_\odot \,.
\end{equation} \\

Combining this equation for maximum soliton mass with the $M-M_{\text{halo}}$ relation (equation \ref{solitonhalo}) we can show that the maximum halo mass is independent of FDM particle mass and takes a value of $(M_{\text{halo}})_{\, \text{max}} = 3.10 \times 10^{20} M_{\odot}$. The combinations of particle and halo mass restricted by the minimum soliton mass are included on Figure~\ref{fig:acctime}. \\

\subsection{Soliton (re-)condensation timescale}
\label{sec:recond}

During structure formation in a cosmological context, solitons rapidly condense in newly-formed halos on the halo free-fall timescale \citep{schive2014cosmic}, which is given by:

\begin{equation}\label{freefalltime}
    t_{\text{con}} = \sqrt{\frac{3 \pi}{32 G \rho}} \,.
\end{equation}

where $\rho$ is the density of the halo. For any halo, we can roughly estimate $\rho$ in equation~\ref{freefalltime} as $\rho_{200} \equiv 200 \rho_{\text{crit}}$ where $\rho_{\text{crit}}$ is the critical density of the Universe, which gives a condensation time $t_{\text{con}} \sim 10^{9}$~yr. \\

It is possible that solitons accreted by SMBHs may re-condense.
This happens on a longer timescale, the kinetic timescale given by \cite{levkov2018gravitational}, because the halo (virialized) is now supported against gravitational collapse by velocity dispersion.
This amounts to a time longer than the free-fall time by a factor that scales as the cube of ratio between halo radius and de Broglie wavelength.
Namely, the timescale can be approximated as:
\begin{equation}\label{cond2}
    t_{\text{re-con}} \simeq \frac{\sqrt{2}b}{12\pi^3} \left(\frac{m}{\hbar}\right)^3 \frac{\sigma^6}{G^2\rho_{\rm c}^2\ln\left( R m\sigma/\hbar\right)} \, ,
\end{equation}
where $R$ is the halo size, $\sigma$ is the velocity dispersion in the halo, and $\rho_{\rm c}$ is the central density of the halo. $b$ is an order unity parameter, where simulations find $b\simeq 0.1$ \citep{eggemeier2019formation}. For $m\gtrsim 10^{-21}$~eV, this timescale can be longer than the age of the Universe for Milky Way or M87-like halos and so re-condensation is not expected to occur for these systems. \\




\begin{figure}
    \centering
    \includegraphics[width=0.47\textwidth]{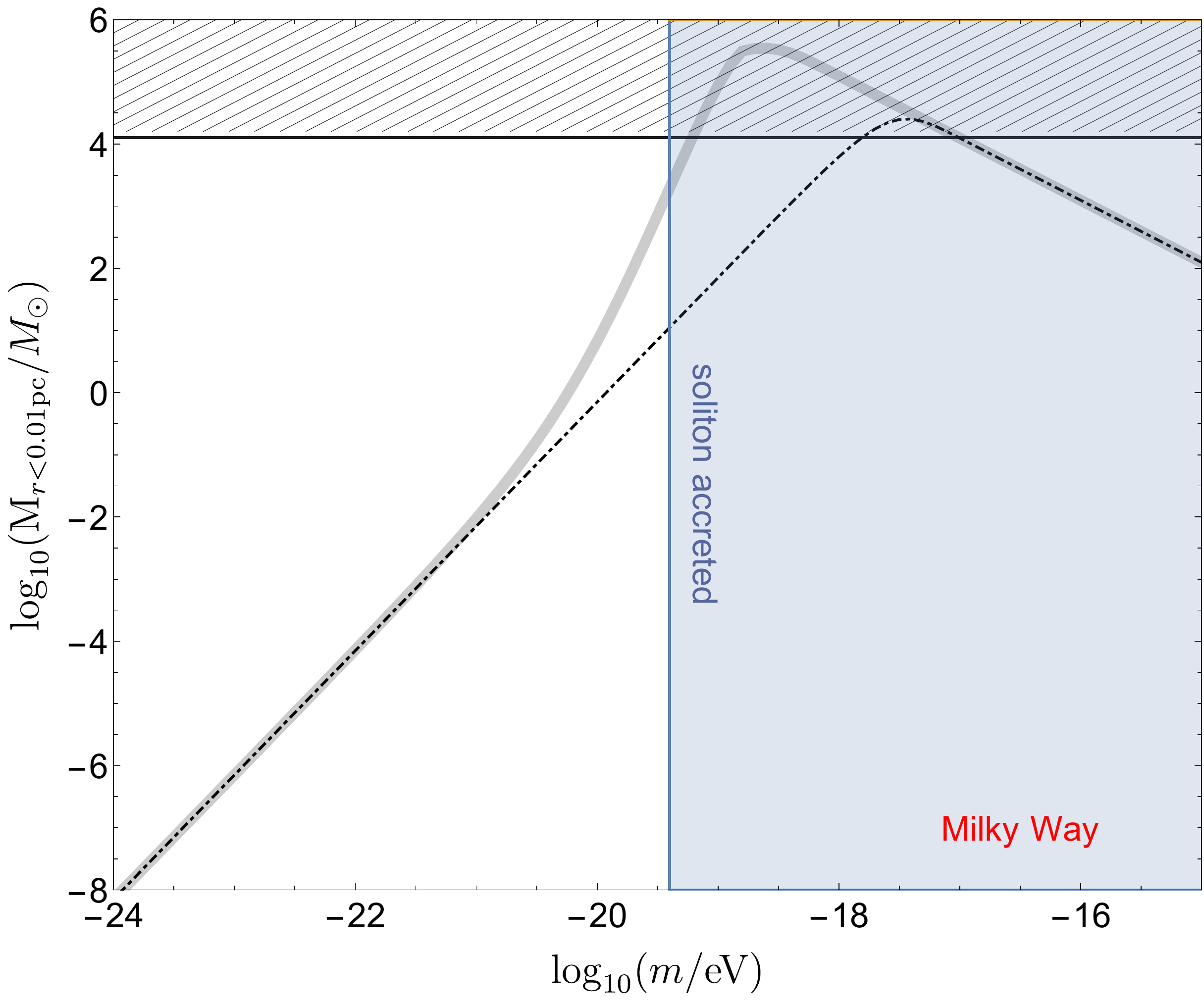}
    \caption{The relationship between enclosed soliton mass and FDM particle mass for the Milky Way, both without the black hole perturber (dotted black line) and with the black hole perturber (solid gray line). The shaded blue region is the estimated range of FDM particle masses that result in the soliton being accreted, which was determined by the selected timescale of $10^{10}$ years. The hatched black region represents the excluded soliton masses for the Milky Way, given by our maximum mass constraint $M_{\text{max}}(r \leq 0.01\text{pc}) \approx 10^4~\si{M_{\odot}}$.}
    \label{fig:figMW}
\end{figure}

\begin{figure}
    \centering
    \includegraphics[width=0.47\textwidth]{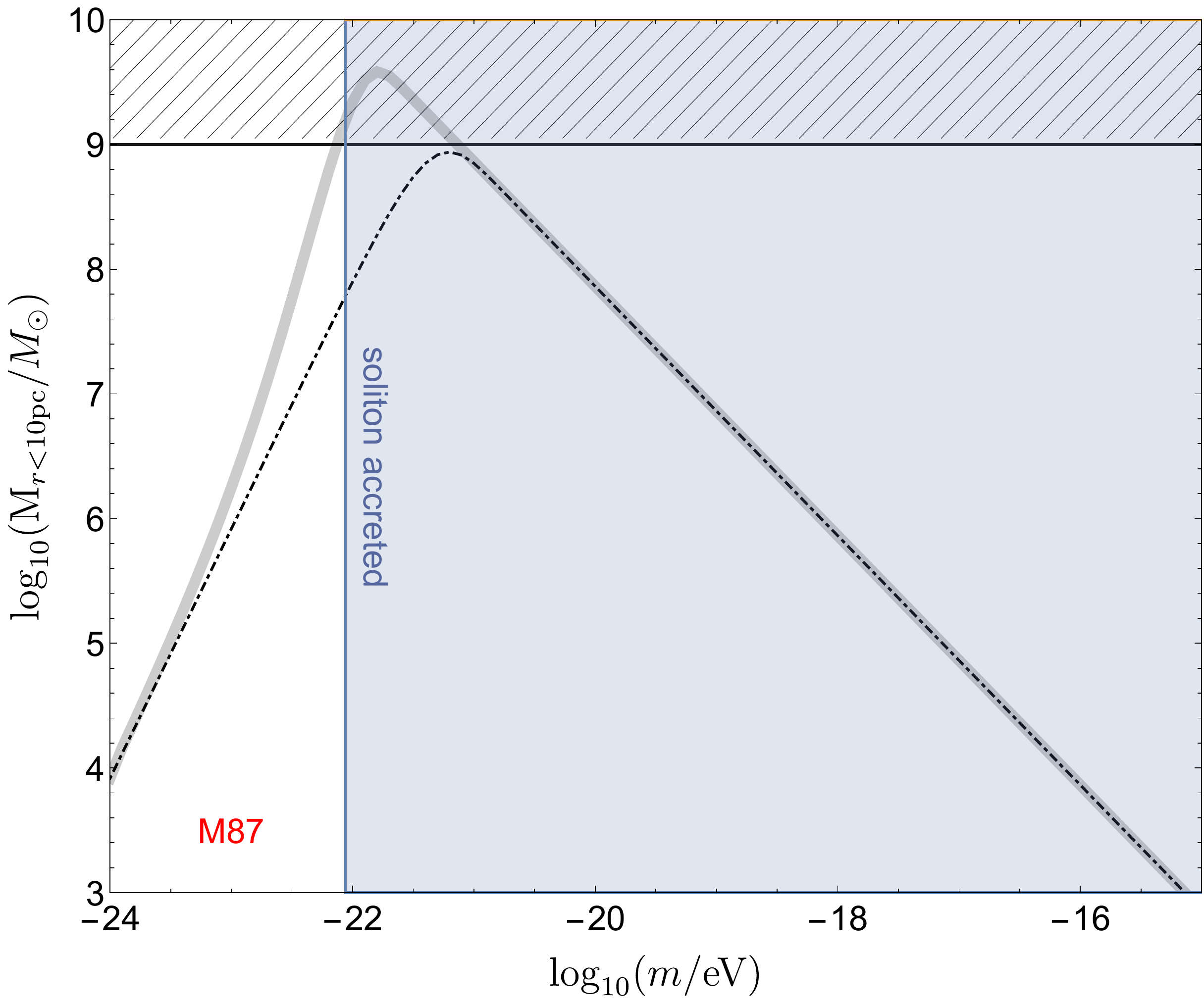}
    \caption{The relationship between enclosed soliton mass and FDM particle mass for the galaxy M87, both without the black hole perturber (dotted black line) and with the black hole perturber (solid gray line). The shaded blue region is the estimated range of FDM particle masses that result in the soliton being accreted, which was determined by the selected timescale of $10^{10}$ years. The hatched black region represents the excluded soliton masses for M87, given by our maximum mass constraint $M_{\text{max}}(r \leq 10\text{pc}) = 10^9~\si{M_{\odot}}$}
    \label{fig:figM87}
\end{figure}

\subsection{Discussion on assumption on soliton mass in the presence of a SMBH} \label{sec:assump}

We have assumed that the soliton mass in the halo is unchanged by the presence of the SMBH, 
and that the soliton-halo mass relation of equation~\ref{solitonhalo} still holds.
This may be a reasonable assumption if the soliton and halo form early in the history of the Universe on a free-fall time (as seen in cosmological simulations) and later the SMBH grows to be significant and traps the soliton in its sphere of influence.
But we point out that this assumption is the main limiting assumption of the work, because a reduced soliton mass would weaken our FDM particle mass constraints. \\

In order to verify that the soliton mass is unchanged when we add a massive black hole at the center, we have carried out numerical simulations of idealized 3D virialized halos with a growing point mass potential at the center. We extended the simulations of \cite{mocz2017galaxy} by growing a massive SMBH at the centre of the halo, as presented in Appendix~\ref{appendSim}.
The profile for the final soliton with the presence of the SMBH ($\Xi=1$) approaches the exponential black hole Bohr radius with normalization set by our assumption of unchanged mass (Figure~\ref{fig:sim}). Mass is conserved to within $1$~per~cent. \\

We discuss an alternative assumption where the soliton mass may be reduced. In this scenario the coupled halo--soliton--SMBH system may thermodynamically re-establish the mass of the soliton once the SMBH grows to be significantly larger in mass than the soliton core. 
In this case, from the potential theory argument,
$[GM/R]_{\rm soliton, no SMBH}\sim [GM/R_{\rm soliton, w/ SMBH}]$.
The limit $\Xi > 1$ would be affected compared to our previous assumption of 
constant mass ($[M]_{\rm soliton, no SMBH}= [M_{\rm soliton, w/ SMBH}]$) 
because the soliton size shrinks significantly with the presence of the black hole.
Since the soliton radius must scale inversely with $M_\bullet$ in the limit $\Xi > 1$, the soliton mass must decrease by a factor of $\Xi$ under this new set of assumptions.
There may be additional constant factors of order unity involved related to the geometric pre-factors of the precise calculation of the potential energy of the soliton, which changes its concentration in the presence of a black hole.
Whether this alternative scenario can be realized in a cosmological setting remains to be seen in full-physics cosmological simulations, and depends on how `conduction' of heat in FDM operates.
We discuss the impact of this alternative assumption on our constraints on the FDM particle mass in section~\ref{section4}.
However, we do point out that our simulation in Appendix~\ref{appendSim} verifies our analytic theory, with the central soliton not changing its mass due to its embedding in a fluctuating halo. \\

\section{FDM mass constraints from the Milky Way and M87}\label{section4}

Observations of stellar dynamics place constraints on the total enclosed mass within central regions of galaxies, including the SMBH mass as well as upper limits on the amount of dark matter within a fixed radius. Thus we can place constraints on the FDM particle mass by testing which values of $m$ produce central soliton profiles consistent with these observational results. \\

First, we consider the Milky Way, which is known to host a SMBH,  Sgr A$^*$, with mass $M_{\bullet} = 4.02(\pm 0.2) \times 10^6 ~\si{M_{\odot}}$  obtained from precise measurements and orbital modeling of short-period stars around the black hole \citep{boehle2016improved,do2019relativistic}.
Observations of the star S0-2 also allow one to constrain the amount of dark matter contained within a radius of $0.011$~pc around Sgr A$^*$ to less than $12.7 \times 10^3 M_{\odot}$ at the 95\% confidence level, independent of the dark matter profile shape \citep{do2019relativistic}.
We considered that the Milky Way resides in a dark matter halo of mass $M_{\text{halo}} = 10^{12}~\si{M_{\odot}}$ \citep{hui2017ultralight} to predict the corresponding soliton mass.
We compute the predicted enclosed dark matter mass within $0.01$~pc of the SMBH as a function of FDM particle mass, and show the results in Figure~\ref{fig:figMW}.
At low particle masses, the core is much larger than the enclosed region and only a small fraction of the soliton mass is contained within the radius. But with increasing FDM particle mass, more dark matter mass is contained in the enclosing region, until the entire soliton fits inside it at $m\simeq 10^{-18.5}~{\rm eV}$, after which the enclosed dark matter mass decreases with particle mass because the total soliton mass decreases with $m$. The presence of Sgr A$^*$ is seen to significantly squeeze the soliton and concentrate it within $0.01$~pc of the SMBH for mass range $ -19.5 \lesssim \log_{10}(m) \lesssim -18$, to the point where it would violate observational limits (black hatched region).
However, we also find that the for $m \gtrsim 10^{-19.4}~{\rm eV}$ the soliton would not survive but would have been accreted by the black hole. The blue shaded region in the plot shows the particle masses over which $t_{\text{acc}} < 10^{10}~{\rm yr}$, i.e., for which axion masses the soliton would be accreted.
Thus we are unable to place a constraint on the particle mass using the Sgr A*. 
But improved longer baseline dynamical mass measurements of the enclosed dark matter that one obtains from measurements of precession of the star S0-2 orbiting the Galactic Center supermassive black hole \citep{boehle2016improved,do2019relativistic} will be useful to place constraints on the FDM particle mass below $10^{-19.4}~{\rm eV}$ in the near future. \\

Second, we consider M87, a supergiant elliptical galaxy in the Virgo cluster.
Recently, the Event Horizon Telescope (EHT) created the first-ever image of a SMBH that resides at the centre of M87 \citep{event2019first}.
The black hole mass was found to be $6.5 \: (\pm0.5) \times 10^9~\si{M_{\odot}}$ based on the size of the black hole shadow / emission ring in the image.
Previously, stellar velocity dispersion measurements in the bulge of M87 estimated that a total of $6.6 \: (\pm0.4) \times 10^9~\si{M_{\odot}}$ of material is contained within $17$--$25$ pc of the centre \citep{gebhardt2011black}. The EHT image confirms that the majority of this mass is the SMBH (and not a supposed FDM soliton).
This leaves little room for a soliton core to be present in the centre of the halo as well.
Even less if one instead considers the total enclosed mass in the central region estimated from gas dynamical measurements: $3.5 \: (\pm0.8) \times 10^9~\si{M_{\odot}}$ \citep{walsh2013m87}, which is systematically less that the stellar velocity dispersion derived result and the EHT image by a factor of $2$. For a conservative measure, we take that a soliton mass of only around $10^9~\si{M_{\odot}}$ ($1/6$ of the black hole mass) can be contained within $10$~pc of the SMBH.
 For M87, we suppose a halo mass of $M_{\text{halo}} = 2 \times 10^{14}~\si{M_{\odot}}$ \citep{hui2017ultralight} to estimate the soliton mass.
 We then calculate the enclosed dark matter mass as a function of particle mass as before.
 Figure~\ref{fig:figM87} summarises the results of our calculations. The soliton core is not expected to survive for particle masses $m>10^{-22.06}~{\rm eV}$. But the mass range $ -22.12 \lesssim \log_{10}(m) \lesssim -22.06$ predicts a soliton core concentrated around the black hole that violates observations. \\
 Again, the lower limit can be extended to lower masses with improved observational measurements on the centrally-concentrated total enclosed mass in the system.
 
Our conclusions for the FDM particle mass constraints are not significantly changed if we assume the alternate scaling of soliton mass with halo mass of section~\ref{sec:assump} in the limit of $\Xi> 1$. This is because $\Xi$ is still close to order unity in the restricted particle mass range and the change of assumptions reduce the enclosed mass only by a factor of $\Xi$.
E.g., for M87 we have $0.71<\Xi<0.78$ for $-22.12 \lesssim \log_{10}(m) \lesssim -22.06$. 
We have also numerically verified our assumptions with simulations for $\Xi=1$ (Appendix~\ref{appendSim}).
\\

\section{Concluding remarks}\label{section5}

We calculated density profiles of soliton cores in the centre of FDM halos by solving the SP equations with, and without, the addition of a black hole point mass perturber. Without perturbation there is a single, universal soliton profile.
With the perturber, the profile depends on the mass ratio of the black hole to soliton: $\Xi\equiv M_{\bullet}/M$. It can be clearly seen in Figure~\ref{fig:haloprofiles} that for sufficiently large halo masses the effect of a black hole point mass perturber in the SP equations is to `squeeze' the soliton inwards, increasing the central density and reducing the core radius. In this regime the size of the core radius scales inversely with black hole mass rather than the soliton mass. This squeezing effect of the soliton is enhanced when the FDM particle mass $m$ is increased because the predicted soliton mass in the halo is decreased, boosting $\Xi$. We showed the degree to which this effect manifests is more evident at higher halo masses for a given $m$.  \\

We applied our calculations to place new constraints on the FDM particle mass given observational constraints on upper limits to the amount of dark matter concentrated near the SMBHs of the Milky Way and M87. 
In the Milky Way, we found that for certain FDM particle masses the FDM soliton core would concentrate around the black hole so much that it would violate the observational mass constraints. However, taking into account the accretion time of the soliton by its SMBH, we actually predict the soliton would not exist in the mass range of $m>10^{-19.4}~{\rm eV}$, so we are not able to constrain the particle mass with given the observational constraints (Figure~\ref{fig:figMW}).
 We were, however, successful in constraining $m$ using M87 (Figure~\ref{fig:figM87}). Based on M87 alone, our conclusions are that the range $ -22.12 \lesssim \log_{10}(m) \lesssim -22.06$ should be forbidden. This result is quite striking as it rules out some of the range of mass right around the original astrophysically-motivated mass of $m\sim10^{-22}~{\rm eV}$ for FDM. Additionally, this constraint would be widened if one were to consider a less conservative estimate for the upper limit on halo mass. Our analysis, namely treating the effect on the soliton of a SMBH perturber, was crucial to obtain this constraint, as an unperturbed soliton is consistent with current observational limits. \\
 
 Improved future measurements of the black hole mass and the total enclosed mass in M87 and the Milky Way will 
 help constrain more of the FDM particle mass parameter space: placing constraints at $m<10^{-22.12}~{\rm eV}$ for M87 and
 $m<10^{-19.4}~{\rm eV}$ for the Milky Way.
 There is also further need for self-consistent time-dependent simulations of a black hole--soliton--halo system, 
 which may increase estimates of the soliton lifetime due to a stabilizing effect of the soliton mass growing larger and thus being accreted slower as it loses mass \citep{du2018tidal}.
 This would place tighter constraints on the FDM particle mass above $m>10^{-22.06}~{\rm eV}$ from M87 and
 $m>10^{-19}~{\rm eV}$ from the Milky Way.
 
 Our mass constraints are generally compatible with other recent astrophysical constraints of $m$ found in the literature, allowing for both small values $m<10^{-22}~{\rm eV}$ preferred by the modeling of dwarf spheroidal galaxies and larger values $m>{\rm few}\times 10^{-22}~\si{eV}$ supported by other considerations. However, our findings are mutually exclusive with the constraints of \cite{broadhurst2019ghostly},  who modeled the dark matter dominated galaxy Antlia II with a FDM soliton core, placing some tension in the FDM theory. \\

\section*{Acknowledgements}

The authors would like to thank Jerry Ostriker and Lachlan Lancaster for reading an earlier version of this manuscript.
E.Y.D. was supported by the 2019 Undergraduate Summer Research Program (USRP) at the Department of Astrophysical Sciences and the Office of Undergraduate Research (OUR) at Princeton University.
P.M. acknowledges support provided by NASA through Einstein Postdoctoral Fellowship grant number PF7-180164 awarded by the Chandra X-ray Center, which is operated by the Smithsonian Astrophysical Observatory for NASA under contract NAS8-03060.



\bibliographystyle{mnras}
\bibliography{bibliog}

\begin{thebibliography}{}
\makeatletter
\relax
\def\mn@urlcharsother{\let\do\@makeother \do\$\do\&\do\#\do\^\do\_\do\%\do\~}
\def\mn@doi{\begingroup\mn@urlcharsother \@ifnextchar [ {\mn@doi@}
  {\mn@doi@[]}}
\def\mn@doi@[#1]#2{\def\@tempa{#1}\ifx\@tempa\@empty \href
  {http://dx.doi.org/#2} {doi:#2}\else \href {http://dx.doi.org/#2} {#1}\fi
  \endgroup}
\def\mn@eprint#1#2{\mn@eprint@#1:#2::\@nil}
\def\mn@eprint@arXiv#1{\href {http://arxiv.org/abs/#1} {{\tt arXiv:#1}}}
\def\mn@eprint@dblp#1{\href {http://dblp.uni-trier.de/rec/bibtex/#1.xml}
  {dblp:#1}}
\def\mn@eprint@#1:#2:#3:#4\@nil{\def\@tempa {#1}\def\@tempb {#2}\def\@tempc
  {#3}\ifx \@tempc \@empty \let \@tempc \@tempb \let \@tempb \@tempa \fi \ifx
  \@tempb \@empty \def\@tempb {arXiv}\fi \@ifundefined
  {mn@eprint@\@tempb}{\@tempb:\@tempc}{\expandafter \expandafter \csname
  mn@eprint@\@tempb\endcsname \expandafter{\@tempc}}}

\bibitem[\protect\citeauthoryear{Amorisco \& Loeb}{Amorisco \&
  Loeb}{2018}]{amorisco2018first}
Amorisco N.~C.,  Loeb A.,  2018, arXiv preprint arXiv:1808.00464

\bibitem[\protect\citeauthoryear{Armengaud, Palanque-Delabrouille, Y{\`e}che,
  Marsh  \& Baur}{Armengaud et~al.}{2017}]{armengaud2017constraining}
Armengaud E.,  Palanque-Delabrouille N.,  Y{\`e}che C.,  Marsh D.~J.,   Baur
  J.,  2017, Monthly Notices of the Royal Astronomical Society, 471, 4606

\bibitem[\protect\citeauthoryear{Avilez, Bernal, Padilla  \& Matos}{Avilez
  et~al.}{2018}]{avilez2018possibility}
Avilez A.~A.,  Bernal T.,  Padilla L.~E.,   Matos T.,  2018, Monthly Notices of
  the Royal Astronomical Society, 477, 3257

\bibitem[\protect\citeauthoryear{Bahrami, Gro{\ss}ardt, Donadi  \&
  Bassi}{Bahrami et~al.}{2014}]{bahrami2014schrodinger}
Bahrami M.,  Gro{\ss}ardt A.,  Donadi S.,   Bassi A.,  2014, New Journal of
  Physics, 16, 115007

\bibitem[\protect\citeauthoryear{Bandara, Crampton  \& Simard}{Bandara
  et~al.}{2009}]{bandara2009relationship}
Bandara K.,  Crampton D.,   Simard L.,  2009, The Astrophysical Journal, 704,
  1135

\bibitem[\protect\citeauthoryear{Bar, Blum, Lacroix  \& Panci}{Bar
  et~al.}{2019}]{bar2019looking}
Bar N.,  Blum K.,  Lacroix T.,   Panci P.,  2019, arXiv preprint
  arXiv:1905.11745

\bibitem[\protect\citeauthoryear{Barranco, Bernal, Degollado, Diez-Tejedor,
  Megevand, Alcubierre, N{\'u}{\~n}ez  \& Sarbach}{Barranco
  et~al.}{2011}]{barranco2011black}
Barranco J.,  Bernal A.,  Degollado J.~C.,  Diez-Tejedor A.,  Megevand M.,
  Alcubierre M.,  N{\'u}{\~n}ez D.,   Sarbach O.,  2011, Physical Review D, 84,
  083008

\bibitem[\protect\citeauthoryear{Barranco, Bernal, Degollado, Diez-Tejedor,
  Megevand, Alcubierre, N{\'u}{\~n}ez  \& Sarbach}{Barranco
  et~al.}{2012}]{barranco2012schwarzschild}
Barranco J.,  Bernal A.,  Degollado J.~C.,  Diez-Tejedor A.,  Megevand M.,
  Alcubierre M.,  N{\'u}{\~n}ez D.,   Sarbach O.,  2012, Physical review
  letters, 109, 081102

\bibitem[\protect\citeauthoryear{Barranco, Bernal, Degollado, Diez-Tejedor,
  Megevand, N{\'u}{\~n}ez  \& Sarbach}{Barranco
  et~al.}{2017}]{barranco2017self}
Barranco J.,  Bernal A.,  Degollado J.~C.,  Diez-Tejedor A.,  Megevand M.,
  N{\'u}{\~n}ez D.,   Sarbach O.,  2017, Physical Review D, 96, 024049

\bibitem[\protect\citeauthoryear{Binney \& Tremaine}{Binney \&
  Tremaine}{2011}]{binney2011galactic}
Binney J.,  Tremaine S.,  2011, Galactic dynamics.
 Vol. 20, Princeton university press

\bibitem[\protect\citeauthoryear{Boehle et~al.,}{Boehle
  et~al.}{2016}]{boehle2016improved}
Boehle A.,  et~al., 2016, The Astrophysical Journal, 830, 17

\bibitem[\protect\citeauthoryear{Bolton, Burles, Koopmans, Treu  \&
  Moustakas}{Bolton et~al.}{2006}]{bolton2006sloan}
Bolton A.~S.,  Burles S.,  Koopmans L.~V.,  Treu T.,   Moustakas L.~A.,  2006,
  The Astrophysical Journal, 638, 703

\bibitem[\protect\citeauthoryear{Boylan-Kolchin, Bullock  \&
  Kaplinghat}{Boylan-Kolchin et~al.}{2011}]{boylan2011too}
Boylan-Kolchin M.,  Bullock J.~S.,   Kaplinghat M.,  2011, Monthly Notices of
  the Royal Astronomical Society: Letters, 415, L40

\bibitem[\protect\citeauthoryear{Broadhurst, De~Martino, Luu, Smoot  \&
  Tye}{Broadhurst et~al.}{2019}]{broadhurst2019ghostly}
Broadhurst T.,  De~Martino I.,  Luu H.~N.,  Smoot G.~F.,   Tye S.-H.~H.,  2019,
  arXiv preprint arXiv:1902.10488

\bibitem[\protect\citeauthoryear{Bullock \& Boylan-Kolchin}{Bullock \&
  Boylan-Kolchin}{2017}]{bullock2017small}
Bullock J.~S.,  Boylan-Kolchin M.,  2017, Annual Review of Astronomy and
  Astrophysics, 55, 343

\bibitem[\protect\citeauthoryear{Cardoso, Dias, Hartnett, Middleton, Pani  \&
  Santos}{Cardoso et~al.}{2018}]{cardoso2018constraining}
Cardoso V.,  Dias {\'O}.~J.,  Hartnett G.~S.,  Middleton M.,  Pani P.,   Santos
  J.~E.,  2018, Journal of Cosmology and Astroparticle Physics, 2018, 043

\bibitem[\protect\citeauthoryear{Chavanis}{Chavanis}{2019}]{chavanis2019mass}
Chavanis P.-H.,  2019, The European Physical Journal Plus, 134, 352

\bibitem[\protect\citeauthoryear{Church, Mocz  \& Ostriker}{Church
  et~al.}{2019}]{church2019heating}
Church B.~V.,  Mocz P.,   Ostriker J.~P.,  2019, Monthly Notices of the Royal
  Astronomical Society, 485, 2861

\bibitem[\protect\citeauthoryear{Col{\'\i}n, Avila-Reese, Roca-F{\`a}brega  \&
  Valenzuela}{Col{\'\i}n et~al.}{2016}]{colin2016cosmological}
Col{\'\i}n P.,  Avila-Reese V.,  Roca-F{\`a}brega S.,   Valenzuela O.,  2016,
  The Astrophysical Journal, 829, 98

\bibitem[\protect\citeauthoryear{Collaboration et~al.}{Collaboration
  et~al.}{2019}]{event2019first}
Collaboration E. H.~T.,  et~al., 2019, arXiv preprint arXiv:1906.11238

\bibitem[\protect\citeauthoryear{{De Martino}, {Broadhurst}, {Tye}, {Chiueh}
  \& {Schive}}{{De Martino} et~al.}{2018}]{2018arXiv180708153D}
{De Martino} I.,  {Broadhurst} T.,  {Tye} S. H.~H.,  {Chiueh} T.,   {Schive}
  H.-Y.,  2018, arXiv e-prints, \href
  {https://ui.adsabs.harvard.edu/abs/2018arXiv180708153D} {p. arXiv:1807.08153}

\bibitem[\protect\citeauthoryear{Desjacques \& Nusser}{Desjacques \&
  Nusser}{2019}]{desjacques2019axion}
Desjacques V.,  Nusser A.,  2019, arXiv preprint arXiv:1905.03450

\bibitem[\protect\citeauthoryear{Do et~al.,}{Do
  et~al.}{2019}]{do2019relativistic}
Do T.,  et~al., 2019, Science, 365, 664

\bibitem[\protect\citeauthoryear{Du, Schwabe, Niemeyer  \& B{\"u}rger}{Du
  et~al.}{2018}]{du2018tidal}
Du X.,  Schwabe B.,  Niemeyer J.~C.,   B{\"u}rger D.,  2018, Physical Review D,
  97, 063507

\bibitem[\protect\citeauthoryear{Eggemeier \& Niemeyer}{Eggemeier \&
  Niemeyer}{2019}]{eggemeier2019formation}
Eggemeier B.,  Niemeyer J.~C.,  2019, arXiv preprint arXiv:1906.01348

\bibitem[\protect\citeauthoryear{Feix, Frank, Pargner, Reischke, Schaefer  \&
  Schwetz}{Feix et~al.}{2019}]{feix2019isocurvature}
Feix M.,  Frank J.,  Pargner A.,  Reischke R.,  Schaefer B.~M.,   Schwetz T.,
  2019, Journal of Cosmology and Astroparticle Physics, 2019, 021

\bibitem[\protect\citeauthoryear{Ferrarese}{Ferrarese}{2002}]{ferrarese2002beyond}
Ferrarese L.,  2002, The Astrophysical Journal, 578, 90

\bibitem[\protect\citeauthoryear{Ferrarese \& Merritt}{Ferrarese \&
  Merritt}{2000}]{ferrarese2000fundamental}
Ferrarese L.,  Merritt D.,  2000, The Astrophysical Journal Letters, 539, L9

\bibitem[\protect\citeauthoryear{Flores \& Primack}{Flores \&
  Primack}{1994}]{flores1994observational}
Flores R.~A.,  Primack J.~R.,  1994, arXiv preprint astro-ph/9402004

\bibitem[\protect\citeauthoryear{Gebhardt et~al.,}{Gebhardt
  et~al.}{2000}]{gebhardt2000relationship}
Gebhardt K.,  et~al., 2000, The Astrophysical Journal Letters, 539, L13

\bibitem[\protect\citeauthoryear{Gebhardt, Adams, Richstone, Lauer, Faber,
  G{\"u}ltekin, Murphy  \& Tremaine}{Gebhardt et~al.}{2011}]{gebhardt2011black}
Gebhardt K.,  Adams J.,  Richstone D.,  Lauer T.~R.,  Faber S.,  G{\"u}ltekin
  K.,  Murphy J.,   Tremaine S.,  2011, The Astrophysical Journal, 729, 119

\bibitem[\protect\citeauthoryear{Gonz{\'a}lez-Morales, Marsh, Pe{\~n}arrubia
  \& Ure{\~n}a-L{\'o}pez}{Gonz{\'a}lez-Morales
  et~al.}{2017}]{gonzalez2017unbiased}
Gonz{\'a}lez-Morales A.~X.,  Marsh D.~J.,  Pe{\~n}arrubia J.,
  Ure{\~n}a-L{\'o}pez L.~A.,  2017, Monthly Notices of the Royal Astronomical
  Society, 472, 1346

\bibitem[\protect\citeauthoryear{Goodman}{Goodman}{2000}]{goodman2000repulsive}
Goodman J.,  2000, New Astronomy, 5, 103

\bibitem[\protect\citeauthoryear{Graham \& Driver}{Graham \&
  Driver}{2007}]{graham2007log}
Graham A.~W.,  Driver S.~P.,  2007, The Astrophysical Journal, 655, 77

\bibitem[\protect\citeauthoryear{G{\"u}ltekin et~al.,}{G{\"u}ltekin
  et~al.}{2009}]{gultekin2009m}
G{\"u}ltekin K.,  et~al., 2009, The Astrophysical Journal, 698, 198

\bibitem[\protect\citeauthoryear{Guzman \& Urena-L{\'o}pez}{Guzman \&
  Urena-L{\'o}pez}{2004}]{guzman2004evolution}
Guzman F.~S.,  Urena-L{\'o}pez L.~A.,  2004, Physical Review D, 69, 124033

\bibitem[\protect\citeauthoryear{Hlo{\v{z}}ek, Marsh  \& Grin}{Hlo{\v{z}}ek
  et~al.}{2018}]{hlovzek2018using}
Hlo{\v{z}}ek R.,  Marsh D.~J.,   Grin D.,  2018, Monthly Notices of the Royal
  Astronomical Society, 476, 3063

\bibitem[\protect\citeauthoryear{Hu, Barkana  \& Gruzinov}{Hu
  et~al.}{2000}]{hu2000fuzzy}
Hu W.,  Barkana R.,   Gruzinov A.,  2000, Physical Review Letters, 85, 1158

\bibitem[\protect\citeauthoryear{Hui, Ostriker, Tremaine  \& Witten}{Hui
  et~al.}{2017}]{hui2017ultralight}
Hui L.,  Ostriker J.~P.,  Tremaine S.,   Witten E.,  2017, Physical Review D,
  95, 043541

\bibitem[\protect\citeauthoryear{Hui, Kabat, Li, Santoni  \& Wong}{Hui
  et~al.}{2019}]{hui2019black}
Hui L.,  Kabat D.,  Li X.,  Santoni L.,   Wong S.~S.,  2019, Journal of
  Cosmology and Astroparticle Physics, 2019, 038

\bibitem[\protect\citeauthoryear{Ir{\v{s}}i{\v{c}}, Viel, Haehnelt, Bolton  \&
  Becker}{Ir{\v{s}}i{\v{c}} et~al.}{2017}]{irvsivc2017first}
Ir{\v{s}}i{\v{c}} V.,  Viel M.,  Haehnelt M.~G.,  Bolton J.~S.,   Becker G.~D.,
   2017, Physical review letters, 119, 031302

\bibitem[\protect\citeauthoryear{Klypin, Kravtsov, Valenzuela  \& Prada}{Klypin
  et~al.}{1999}]{klypin1999missing}
Klypin A.,  Kravtsov A.~V.,  Valenzuela O.,   Prada F.,  1999, The
  Astrophysical Journal, 522, 82

\bibitem[\protect\citeauthoryear{Kobayashi, Murgia, De~Simone,
  Ir{\v{s}}i{\v{c}}  \& Viel}{Kobayashi et~al.}{2017}]{kobayashi2017lyman}
Kobayashi T.,  Murgia R.,  De~Simone A.,  Ir{\v{s}}i{\v{c}} V.,   Viel M.,
  2017, Physical Review D, 96, 123514

\bibitem[\protect\citeauthoryear{Kormendy \& Ho}{Kormendy \&
  Ho}{2013}]{kormendy2013coevolution}
Kormendy J.,  Ho L.~C.,  2013, Annual Review of Astronomy and Astrophysics, 51,
  511

\bibitem[\protect\citeauthoryear{{Lazar} \& {Bullock}}{{Lazar} \&
  {Bullock}}{2019}]{2019arXiv190708841L}
{Lazar} A.,  {Bullock} J.~S.,  2019, arXiv e-prints, \href
  {https://ui.adsabs.harvard.edu/abs/2019arXiv190708841L} {p. arXiv:1907.08841}

\bibitem[\protect\citeauthoryear{Levkov, Panin  \& Tkachev}{Levkov
  et~al.}{2018}]{levkov2018gravitational}
Levkov D.,  Panin A.,   Tkachev I.,  2018, Physical review letters, 121, 151301

\bibitem[\protect\citeauthoryear{Lora \& Maga{\~n}a}{Lora \&
  Maga{\~n}a}{2014}]{lora2014sextans}
Lora V.,  Maga{\~n}a J.,  2014, Journal of Cosmology and Astroparticle Physics,
  2014, 011

\bibitem[\protect\citeauthoryear{Ma, Greene, McConnell, Janish, Blakeslee,
  Thomas  \& Murphy}{Ma et~al.}{2014}]{ma2014massive}
Ma C.-P.,  Greene J.~E.,  McConnell N.,  Janish R.,  Blakeslee J.~P.,  Thomas
  J.,   Murphy J.~D.,  2014, The Astrophysical Journal, 795, 158

\bibitem[\protect\citeauthoryear{Macci{\`o}, Paduroiu, Anderhalden, Schneider
  \& Moore}{Macci{\`o} et~al.}{2012}]{maccio2012cores}
Macci{\`o} A.~V.,  Paduroiu S.,  Anderhalden D.,  Schneider A.,   Moore B.,
  2012, Monthly Notices of the Royal Astronomical Society, 424, 1105

\bibitem[\protect\citeauthoryear{Marsh \& Pop}{Marsh \&
  Pop}{2015}]{marsh2015axion}
Marsh D.~J.,  Pop A.-R.,  2015, Monthly Notices of the Royal Astronomical
  Society, 451, 2479

\bibitem[\protect\citeauthoryear{Mocz \& Succi}{Mocz \&
  Succi}{2015}]{mocz2015numerical}
Mocz P.,  Succi S.,  2015, Physical Review E, 91, 053304

\bibitem[\protect\citeauthoryear{Mocz, Vogelsberger, Robles, Zavala,
  Boylan-Kolchin, Fialkov  \& Hernquist}{Mocz et~al.}{2017}]{mocz2017galaxy}
Mocz P.,  Vogelsberger M.,  Robles V.~H.,  Zavala J.,  Boylan-Kolchin M.,
  Fialkov A.,   Hernquist L.,  2017, Monthly Notices of the Royal Astronomical
  Society, 471, 4559

\bibitem[\protect\citeauthoryear{Mocz, Lancaster, Fialkov, Becerra  \&
  Chavanis}{Mocz et~al.}{2018}]{mocz2018schrodinger}
Mocz P.,  Lancaster L.,  Fialkov A.,  Becerra F.,   Chavanis P.-H.,  2018,
  Physical Review D, 97, 083519

\bibitem[\protect\citeauthoryear{Mocz et~al.,}{Mocz
  et~al.}{2019a}]{mocz2019galaxy}
Mocz P.,  et~al., 2019a, arXiv preprint arXiv:1911.05746

\bibitem[\protect\citeauthoryear{Mocz et~al.,}{Mocz
  et~al.}{2019b}]{mocz2019first}
Mocz P.,  et~al., 2019b, Physical review letters, 123, 141301

\bibitem[\protect\citeauthoryear{Moore}{Moore}{1994}]{moore1994evidence}
Moore B.,  1994, Nature, 370, 629

\bibitem[\protect\citeauthoryear{Moore, Ghigna, Governato, Lake, Quinn, Stadel
  \& Tozzi}{Moore et~al.}{1999}]{moore1999dark}
Moore B.,  Ghigna S.,  Governato F.,  Lake G.,  Quinn T.,  Stadel J.,   Tozzi
  P.,  1999, The Astrophysical Journal Letters, 524, L19

\bibitem[\protect\citeauthoryear{Moroz, Penrose  \& Tod}{Moroz
  et~al.}{1998}]{moroz1998spherically}
Moroz I.~M.,  Penrose R.,   Tod P.,  1998, Classical and Quantum Gravity, 15,
  2733

\bibitem[\protect\citeauthoryear{Nadler, Gluscevic, Boddy  \& Wechsler}{Nadler
  et~al.}{2019}]{nadler2019constraints}
Nadler E.~O.,  Gluscevic V.,  Boddy K.~K.,   Wechsler R.~H.,  2019, The
  Astrophysical Journal Letters, 878, L32

\bibitem[\protect\citeauthoryear{Nori, Murgia, Ir{\v{s}}i{\v{c}}, Baldi  \&
  Viel}{Nori et~al.}{2018}]{nori2018lyman}
Nori M.,  Murgia R.,  Ir{\v{s}}i{\v{c}} V.,  Baldi M.,   Viel M.,  2018,
  Monthly Notices of the Royal Astronomical Society, 482, 3227

\bibitem[\protect\citeauthoryear{Safarzadeh \& Spergel}{Safarzadeh \&
  Spergel}{2019}]{safarzadeh2019ultra}
Safarzadeh M.,  Spergel D.~N.,  2019, arXiv preprint arXiv:1906.11848

\bibitem[\protect\citeauthoryear{Sanchez et~al.,}{Sanchez
  et~al.}{2018}]{sanchez2018preferential}
Sanchez N.~N.,  et~al., 2018, The Astrophysical Journal, 860, 20

\bibitem[\protect\citeauthoryear{Schive, Chiueh  \& Broadhurst}{Schive
  et~al.}{2014a}]{schive2014cosmic}
Schive H.-Y.,  Chiueh T.,   Broadhurst T.,  2014a, Nature Physics, 10, 496

\bibitem[\protect\citeauthoryear{Schive, Liao, Woo, Wong, Chiueh, Broadhurst
  \& Hwang}{Schive et~al.}{2014b}]{schive2014understanding}
Schive H.-Y.,  Liao M.-H.,  Woo T.-P.,  Wong S.-K.,  Chiueh T.,  Broadhurst T.,
    Hwang W.~P.,  2014b, Physical review letters, 113, 261302

\bibitem[\protect\citeauthoryear{Soltan}{Soltan}{1982}]{soƚtan1982masses}
Soltan A.,  1982, Monthly Notices of the Royal Astronomical Society, 200, 115

\bibitem[\protect\citeauthoryear{Sun, Greene, Impellizzeri, Kuo, Braatz  \&
  Tuttle}{Sun et~al.}{2013}]{sun2013refining}
Sun A.-L.,  Greene J.~E.,  Impellizzeri C.~V.,  Kuo C.-Y.,  Braatz J.~A.,
  Tuttle S.,  2013, The Astrophysical Journal, 778, 47

\bibitem[\protect\citeauthoryear{Tremaine \& Gunn}{Tremaine \&
  Gunn}{1979}]{tremaine1979dynamical}
Tremaine S.,  Gunn J.~E.,  1979, Physical Review Letters, 42, 407

\bibitem[\protect\citeauthoryear{Unruh}{Unruh}{1976}]{unruh1976absorption}
Unruh W.,  1976, Physical Review D, 14, 3251

\bibitem[\protect\citeauthoryear{Veltmaat, Niemeyer  \& Schwabe}{Veltmaat
  et~al.}{2018}]{veltmaat2018formation}
Veltmaat J.,  Niemeyer J.~C.,   Schwabe B.,  2018, Physical Review D, 98,
  043509

\bibitem[\protect\citeauthoryear{Veltmaat, Schwabe  \& Niemeyer}{Veltmaat
  et~al.}{2019}]{veltmaat2019baryon}
Veltmaat J.,  Schwabe B.,   Niemeyer J.~C.,  2019, arXiv preprint
  arXiv:1911.09614

\bibitem[\protect\citeauthoryear{Walsh, Barth, Ho  \& Sarzi}{Walsh
  et~al.}{2013}]{walsh2013m87}
Walsh J.~L.,  Barth A.~J.,  Ho L.~C.,   Sarzi M.,  2013, The Astrophysical
  Journal, 770, 86

\bibitem[\protect\citeauthoryear{{Wasserman} et~al.,}{{Wasserman}
  et~al.}{2019}]{2019arXiv190510373W}
{Wasserman} A.,  et~al., 2019, arXiv e-prints, \href
  {https://ui.adsabs.harvard.edu/abs/2019arXiv190510373W} {p. arXiv:1905.10373}

\bibitem[\protect\citeauthoryear{Weinberg, Bullock, Governato, de Naray  \&
  Peter}{Weinberg et~al.}{2015}]{weinberg2015cold}
Weinberg D.~H.,  Bullock J.~S.,  Governato F.,  de Naray R.~K.,   Peter A.~H.,
  2015, Proceedings of the National Academy of Sciences, 112, 12249

\makeatother
\end{thebibliography}



\appendix 

\section{Tables}\label{append}
This appendix (Tables~\ref{tab:example} and \ref{tab:example2}) presents the values of $\hat{\alpha}, \: \hat{\gamma}_0$, $\hat{M}$ (along with corresponding halo mass) for the numerically produced solutions which are shown in Figure~\ref{fig:haloprofiles}. The dimensionless mass was calculated by numerical integration of $\hat{\phi}(r)$, illustrated by equation~\ref{dimensionlessmass}. 

\FloatBarrier

\begin{table}
 \caption{Values of $\hat{\alpha}$ chosen to produce density profiles of the relevant halo masses, along with the corresponding $\hat{\gamma}_0$ and $\hat{M}$ for a FDM particle mass of $10^{-22}~\si{eV}.$}
 \label{tab:example}
 \begin{tabular}{lccc}
  \hline
  $\hat{\alpha}$ & $\hat{\gamma}_0$ & $\hat{M}$ & Halo Mass\\
   & & & $M_{\sun}$ \\
  \hline
  $1.58 \times 10^{-7}$ & 0.649535 & 2.06219 & $10^{8}$\\
  $1.58 \times 10^{-6}$ & 0.649533 & 2.06219 & $10^{9}$\\
  $5.01 \times 10^{-5}$ & 0.649478 & 2.06204 & $10^{10}$\\
  $3.98 \times 10^{-4}$ & 0.649082 & 2.0504 & $10^{11}$\\
  $6.31 \times 10^{-3}$ & 0.642371 & 2.04351 & $10^{12}$\\
  0.10 & 0.537464 & 1.77119 & $10^{13}$\\
  0.59 & 0.0251269 & 0.632804 & $10^{14}$\\
  \hline
 \end{tabular}
\end{table}

\begin{table}
 \caption{Values of $\hat{\alpha}$ chosen to produce density profiles of the relevant halo masses, along with the corresponding $\hat{\gamma}_0$ and $\hat{M}$ for a FDM particle mass of $10^{-21}~\si{eV}.$}
 \label{tab:example2}
 \begin{tabular}{lccc}
  \hline
  $\hat{\alpha}$ & $\hat{\gamma}_0$ & $\hat{M}$ & Halo Mass\\
   & & & $M_{\sun}$ \\
  \hline
  $10^{-7}$ & 0.649535 & 2.06219 & $10^{8}$\\
  $1.58 \times 10^{-5}$ & 0.649517 & 2.06219 & $10^{9}$\\
  $2.51 \times 10^{-4}$ & 0.649249 & 2.06145 & $10^{10}$\\
  $3.98 \times 10^{-3}$ & 0.645013 & 2.0504 & $10^{11}$\\
  $0.06$ & 0.581918 & 1.88620 & $10^{12}$\\
  $0.48$ & 0.136695 & 0.830543 & $10^{13}$\\
  $1.21$ & -0.669924 & 0.131330 & $10^{14}$\\
  \hline
 \end{tabular}
\end{table}



\section{Simulation of virialized FDM with SMBH}\label{appendSim}

\begin{figure}
    \centering
    \includegraphics[width=0.47\textwidth]{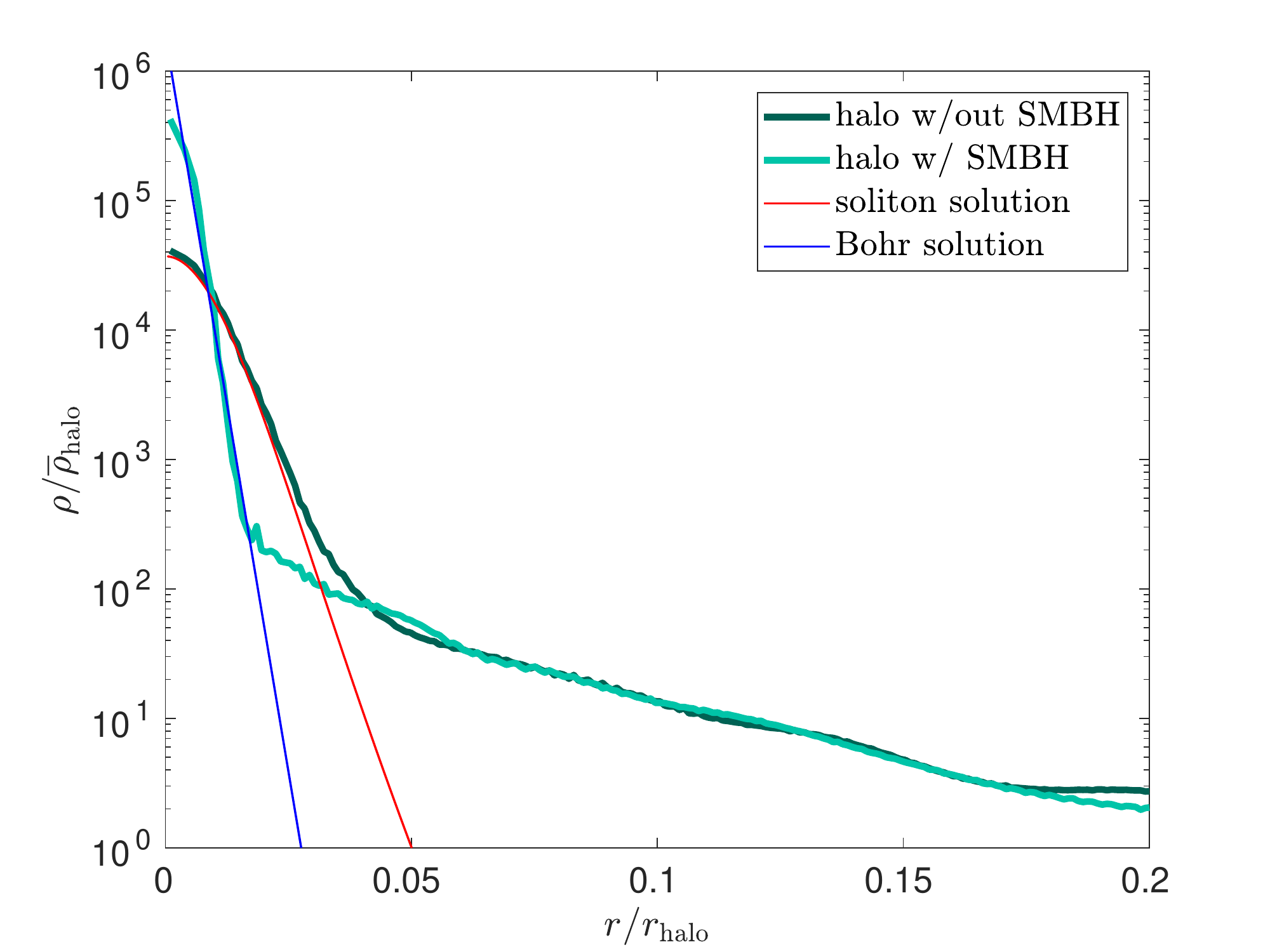}
    \caption{Simulations of a $M_{\rm soliton}/M_{\rm halo}=0.3$ FDM halo with a soliton at its centre (dark green) that gains a SMBH of mass $M_\bullet/M_{\rm halo}=0.3$ and is allowed to relax again (light green). The final radial profile approaches the hydrogen atom solution with radius set by the black hole mass `Bohr radius', and normalization set by assuming mass conservation of the soliton.}
    \label{fig:sim}
\end{figure}

To verify our analytic theory and some of our assumptions, 
we have carried out a numerical simulation 
by placing a SMBH as a point mass into one of our simulations
of an idealized virialized FDM halo from \cite{mocz2017galaxy} (details of numerical implementation are found in cited paper. Simulation resolution is $256^3$).
The system was initially a virialized FDM halo with soliton-to-halo mass ratio of $M/M_{\rm halo}=0.3$, with a virialization timescale of timescale $T_{\rm vir}$.
We then grew a SMBH at the centre of the halo linearly over the timescale $T_{\rm vir}$. The system was allowed to relax again for another period of $T_{\rm vir}$.
Figure~\ref{fig:sim} shows the radial profile of the halo before the addition of the black hole and the final relaxed solution.

\bsp
\label{lastpage}
\end{document}